\documentclass[final,3p,times,twocolumn]{elsarticle}

\usepackage{amssymb}

\usepackage{lineno}
\usepackage{ulem}

\usepackage{booktabs} 
\usepackage[bottom]{footmisc}

\journal{Physical Review D }
\usepackage[dvipsnames]{xcolor}
\usepackage{graphicx}
\usepackage{amsmath}

\begin{document}


\begin{frontmatter}



\title{Cross-section measurement of two-photon annihilation in-flight of positrons at $\sqrt{s}=20$~MeV with the PADME detector}

\author[a]{F.~Bossi}
\author[b]{P.~Branchini}
\author[a]{B.~Buonomo}
\author[c]{V.~Capirossi}
\author[d,e]{A.P.~Caricato}
\author[d]{G.~Chiodini}
\author[a]{R.~De Sangro}
\author[a]{C.~Di Giulio}
\author[a]{D.~Domenici}
\author[f]{F.~Ferrarotto}
\author[a]{G.~Finocchiaro}
\author[a]{L.G~Foggetta}
\author[g]{A.~Frankenthal}
\author[a]{M.~Garattini}
\author[h,i]{G.~Georgiev}
\author[a]{F.~Giacchino\fnref{giacc}}
\author[a]{P.~Gianotti}
\author[i]{S.~Ivanov}
\author[i]{Sv.~Ivanov}
\author[i,a]{V.~Kozhuharov}
\author[f]{E.~Leonardi}
\author[j,f]{E.~Long}
\author[d,e]{M.~Martino}
\author[d,e]{I.~Oceano\corref{cor1}}
\author[d,e]{F.~Oliva\fnref{oliv}}
\author[j,f]{G.C. Organtini}
\author[c]{F.~Pinna}
\author[j]{G.~Piperno}
\author[j,f]{M.~Raggi}
\author[a]{I.~Sarra}
\author[i]{R.~Simeonov}
\author[a]{T.~Spadaro}
\author[d,e]{S.~Spagnolo}
\author[a]{E.~Spiriti}
\author[b]{D.~Tagnani}
\author[a]{C.~Taruggi}
\author[f]{P.~Valente}
\author[a]{E.~Vilucchi}

\address[a]{INFN Laboratori Nazionali di Frascati,  via E. Fermi 54, Frascati, Italy}
\address[b]{INFN sez. Roma 3,  via della vasca navale 84, Roma, Italy}
\address[c]{DISAT Politecnico di Torino and INFN sez. Torino, C.so Duca degli Abruzzi 24, Torino, Italy}
\address[d]{INFN sez. Lecce,  via Provinciale per Arnesano, Lecce, Italy}
\address[e]{Dip. Mat. e Fisica Salento Univ., via Provinciale per Arnesano, Lecce, Italy}
\address[f]{INFN sez. Roma 1, p.le A. Moro 2, Rome, Italy}
\address[g]{Physics Dep. Princeton Univ., Washington Road, Princeton, USA}
\address[h]{INRNE Bulgarian Accademy of Science, 72 Tsarigradsko shosse Blvd., Sofia, Bulgaria }
\address[i]{Sofia Univ. ``St. Kl. Ohridski'', 5 J. Bourchier Blvd., Sofia, Bulgaria}
\address[j]{Dip. di Fisica Sapienza Univ., p.le A. Moro 2, Roma, Italy}


\cortext[cor1]{Corresponding author: isabella.oceano@le.infn.it}
\fntext[giacc]{presently at INFN sez. Roma 2, via della Ricerca Scientifica 1, Rome, Italy}
\fntext[oliv]{presently at School of Physics and Astronomy, Edinburgh Univ., Edinburgh, UK}
\begin{abstract}
The inclusive cross-section of annihilation in flight
$e^+e^-\rightarrow\gamma\gamma$ 
 of 430 MeV positrons with atomic electrons of a thin diamond target has been measured with the PADME detector at the Laboratori Nazionali di Frascati. 
The two photons produced in the process 
were detected by an electromagnetic calorimeter 
made of BGO crystals.
This measurement is the first one based on the direct detection of the photon pair and one of the most precise for positron energies below 1 GeV.
This measurement represents a necessary step to 
search for dark sector particles and mediators weakly coupled to photons and/or electrons with masses ranging from 1 MeV to 20 MeV with PADME. The measurement agrees with the Next to Leading Order QED prediction within the overall 6\% uncertainty.
\end{abstract}



\begin{keyword}



\end{keyword}

\end{frontmatter}

\section{Introduction}
\label{Introduction}

PADME (Positron Annihilation into Dark Matter Experiment) at the Laboratori Nazionali di Frascati (LNF) of INFN is a fixed target experiment exploiting a positron beam. It was designed to search for a hypothetical dark photon $A^{\prime}$ produced in association with a photon in electron-positron annihilation \cite{Raggi_2014}. This particle is postulated to be the gauge boson associated with a $U_d(1)$ symmetry in a sector where dark matter would be confined according to the paradigm of the ``hidden-sector" theoretical models \cite{darksector}. 
A simple model allowing very weak interactions of Standard Model particles with dark matter is obtained from a kinetic mixing~\cite{Fabbrichesi_2021}. 

PADME is expected to be sensitive to the parameter $\epsilon$,  describing the effective coupling between $A^{\prime}$ and the photon, relative to the electromagnetic coupling $\alpha$, for $\epsilon\geq$ 10$^{-3}$ and values of the $A^{\prime}$ mass m$_{A^{\prime}}\le$ 23.7 MeV/c$^{2}$ after collecting $\approx 10^{13}$ Positrons-On-Target (POT) at the energy of  550~MeV. The search technique relies on the reconstruction of the squared missing mass $M^2=(P_{e^+}+P_{e^-}-P_{\gamma})^2$ of single photon final states. The positron four-momentum is determined by the PADME beam-line (see Sec. \ref{subsec:pbeam}). The photon four-momentum $P_{\gamma}$ is measured by a high resolution segmented electromagnetic calorimeter (ECAL), based on the position of the beam spot on the target, where positrons are assumed to annihilate on electrons at rest. 
Therefore, reaching a good understanding of the ECAL performance, through the study of the theoretically well known QED process of electron-positron annihilation in photons $e^+e^-\rightarrow\gamma\gamma$, is a crucial step for the physics goal of the experiment. The measurement of the cross-section of such process allows for 
a calibration of photon reconstruction and a monitor of the beam intensity 
with high precision.

A few measurements of the $e^+e^-\rightarrow\gamma\gamma$ cross-section were performed in the second half of the 1950,  exploiting the disappearance of positron tracks in a target.
Colgate and Gilbert in 1953 provided measurements, with a precision of about 20$\%$, by measuring the attenuation of positrons with energies 50, 100 and 200 MeV \cite{Colgate}. In 1963, Malamud and Weill reached a precision of about 4$\%$ using a bubble chamber and positrons accelerated at the energy of 600~MeV \cite{Malamud}. Fabiani et al. in 1962 at CERN measured the annihilation cross-section for positrons of energies equal to 1.94, 5.80, 7.71,  and 9.64 GeV, with an uncertainty better than about 5$\%$ \cite{Fabiani}.

This paper presents the first direct measurement of the absolute cross-section of annihilation in flight of 430 MeV positrons in photons.
Section \ref{Sec:PADME} describes the PADME beam and detector, along with the photon reconstruction technique. Section \ref{Sec:DataAndSamples} presents the data used for the measurement and the event selection. The measurement strategy is detailed in Section \ref{Sec.:cross-section}  
along with the main uncertainties and the results. The selection requirements are designed to be inclusive of events with extra radiation, therefore the measurement is compared to the QED prediction at the Next-Leading-Order approximation for the inclusive $e^+e^-\rightarrow\gamma\gamma(\gamma^*)$ cross-section. 


\section{The PADME experiment}
\label{Sec:PADME}

\subsection{The positron beam}
\label{subsec:pbeam}
The PADME experiment is located in the BTF (Beam Test Facility) of the Laboratori Nazionali di Frascati where the LINAC of the DA$\Phi$NE collider provides a variable energy positron beam \cite{Buonomo:2015pba,Valente1}.
Before the last bending dipole, a 125 $\mu$m thick mylar window separates the LINAC high vacuum region, from the PADME vacuum region, where less stringent  conditions are required. The data-set used for this analysis was collected from September to November 2020 with a beam energy of about $430~\rm{MeV}$ and an average beam intensity of 25$\times 10^3$ positrons approximately evenly distributed in 250~ns long bunches.

\subsubsection{The diamond active target}
\label{Sec.:Target}
The diamond active target \cite{Oliva:2019alx} allows estimating the particle multiplicity in the bunch and the average position of the beam interaction point. 
It is a full carbon doubled-sided strip detector 100 $\mu$m thick and with an area of $\rm 2 \times 2~cm^2$ made of CVD\footnote{Produced by a Chemical Vapour Deposition process.} poly-crystalline diamond. The pattern of graphitic strip electrodes, obtained by irradiation with an ArF laser, measures coordinates in orthogonal directions on the two sides. 
The thickness of graphite strips was estimated to be about 200 nm and material ablation negligible. Sixteen electrodes per side were connected to front-end electronics.

The target is sitting in the beam pipe vacuum. A remotely controlled step-motor was used to move it out of the beam-line and park it on the side when collecting special data for the study of beam background, i.e. background not originating from interactions of beam particles with the target.


\begin{figure}[htbp]
\begin{center}
\includegraphics[width=\linewidth]{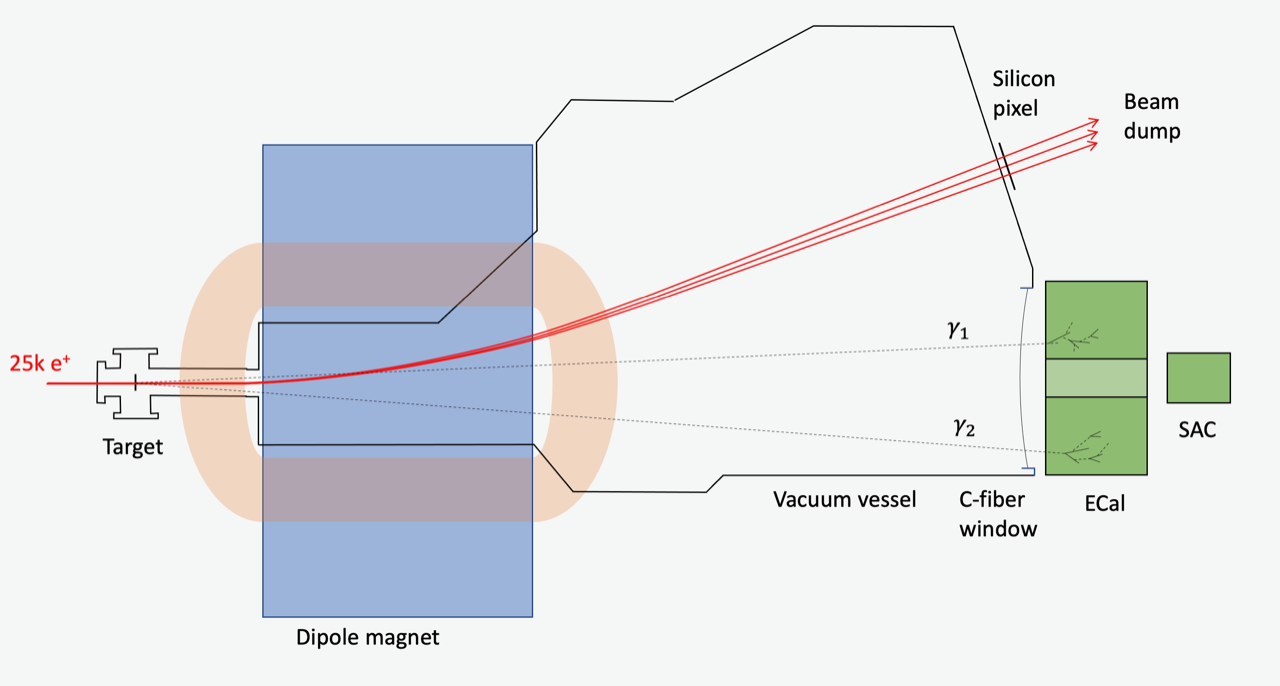}
\caption{The schematic view of the PADME experiment showing the components relevant for the $e^+e^-\rightarrow\gamma\gamma$ cross-section measurement.
}
\label{default}
\end{center}
\end{figure}

\subsection{The electromagnetic calorimeter}
The main detector of the PADME experiment is the electromagnetic calorimeter, ECAL~\cite{EcalArticle}.
It is a segmented calorimeter, made of 616 BGO crystals.
The size of each crystal is 2.1$\ \times \ 2.1\ \times\ \rm 23~cm^3$, to allow the electromagnetic shower to be fully contained in the longitudinal direction and for a fraction equal to 70$\%$ in the transverse direction. The crystals are arranged in a cylindrical array with a central squared hole, corresponding to 5$\times$5 missing crystals, and an external radius of 30 cm.
The central square hole prevents  the calorimeter from being overwhelmed by the high rate of Bremsstrahlung photons emitted in the forward direction. 
The scintillation light is detected by HZC XP1911 type B photomultipliers (19 mm diameter)~\cite{EcalArticle}.
They reach a maximum quantum efficiency of about 21$\%$ at 480 nm, where the light emission of BGO is also maximum. The BGO light emission intensity changes with temperature by about 0.9$\%/^0$C, therefore the ECAL temperature map was monitored with 40 Pt1000 probes.  

At each beam bunch reaching the target, the signals of all photomultipliers were digitized~\cite{Leonardi:2017ocd} in a readout window $1~\mu$s wide at the rate of 1Gs/s with 12bit ADCs. The waveforms were recorded after zero suppression. A cosmic ray trigger was setup to collect data for a continuous monitor of the gain during data taking. These data were also used to estimate the average channel efficiency, which was found to be about 99.6$\%$ for cosmic rays~\cite{Albicocco:2022ukx}. 

\subsection{Photon reconstruction}
\label{Sec.:Reconstruction}
The reconstruction of photons in ECAL was based on a processing on the waveforms allowing for the reconstruction of up to three hits per crystal. The algorithm was based on a signal template derived from data. The energy and time of a hit was estimated from the scale factor and shift of the template required to adjust it to a pulse in the waveform. The template algorithm easily accounted for waveforms with a truncated tail, due to the limited time acquisition window, 
or with a saturated amplitude, due to high energy release or overlapping hits. The hit energies were corrected for relative calibration factors extracted using cosmic rays. An additional absolute calibration factor 
was applied to match the average total energy of annihilation photon pairs to the beam energy.

Hits close in time and space were grouped in clusters not larger than 7$\times 7$  crystals. The clustering procedure starts from a seeding hit of energy equal to at least 20~MeV. All nearby hits with energy above 1~MeV at a distance from the seed not exceeding three crystals and in time coincidence within 6~ns are merged. 
The cluster energy is given by the sum of the energies of all  constituent hits. Transverse position and time are given by energy-weighted averages.

\section{Data sample and event selection}
\label{Sec:DataAndSamples}
\subsection{PADME data and Monte Carlo samples}
The data used for this measurement were collected during the PADME RunII from Sept. to Nov. 2020. The beam energy was stable at the value of 432.5~MeV, with a relative energy spread of 0.5 $\%$ and a typical particle density in the bunch of about 100 POT/ns. 
They correspond to a subset of the RunII data featuring good 
stability in beam spot intensity and position on the target during the run. The main features of the runs are summarised in Tab. \ref{Tab.:Runs}. 
\begin{table}[!ht]
\centering
\caption{Main features of the runs used in this analysis.}
\begin{tabular}{ccccc}
\toprule
Run & NPOT & e$^+$/bunch & bunch length
\\
 $\#$   & $[10^{10}]$ & $[10^3]$ & [ns] 
 \\
\midrule
\midrule
$30369 $ & $  8.2  $ & $27.0\pm1.7$ & $260$     
\\		 
$30386 $ & $  2.8  $ & $19.0\pm1.4$ & $240$     
\\		 
$30547 $ & $  7.1  $ & $31.5\pm1.4$ & $270$     
\\		 
$30553 $ & $  2.8  $ & $35.8\pm1.3$ & $260$    
\\		 
$30563 $ & $  6.0  $ & $26.8\pm1.2$ & $270$     
\\		 
$30617 $ & $  6.1  $ & $27.3\pm1.5$ & $270$     
\\		 
$30624 $ & $  6.6  $ & $29.5\pm2.1$ & $270$      
\\   
$30654 $ & No-target & $\sim 27$    & $\sim 270$ 
\\
$30662 $ & No-Target & $\sim 27$    & $\sim 270$ 
\\        
\bottomrule
\end{tabular}
\label{Tab.:Runs}
\end{table}
The total number of POT, after event quality cuts (see Sec.\ref{Sec.:EventSelection}), amounts to 3.97$\times$10$^{11}$, with an expected yield of two-photon annihilation events of about 5$\times$10$^{5}$. 
Therefore, this data set allows for a measurement with a statistical error smaller than 1\%.

Special runs were collected with the target out of the beam line in order to study in data the background component not related to beam-target interactions that will be referred as beam-background. This component originates from out of orbit positrons and positrons radiating when crossing the mylar vacuum separation window.
These particles can interact with beam line materials and produce showers of secondary particles, sometimes reaching the detectors~\cite{PADME:2022ysa}.


The PADME experiment was simulated in all its components with the GEANT4 software \cite{AGOSTINELLI2003250}.
The positron beam was generated just upstream the target with a bunch multiplicity of 25$\times 10^3$, a Gaussian spot of 1 mm, a divergence of 0.1 mrad and an energy spread of $1 \ \rm{MeV}$.
The dominant QED processes are simulated according to the GEANT4 physics list:  Bremsstrahlung (on nuclei and atomic electrons), annihilation, Bhabha scattering, and other minor effects. 
Therefore, the simulation reproduces the effects of pile-up of interactions of positrons in the target, 
resulting in several photons in the calorimeter overlapping in the same bunch (the so called PADME event),
while it does not describe the beam background\footnote{A simulation of the beam line was also available but it was not used in this work and data driven techniques  were preferred to estimate the background in the analysis.}.

Signal acceptance and event migration induced by resolution are most conveniently estimated from simulations of single annihilation events, free from pile-up and with the true photon kinematics easily accessible.
The CalcHEP generator~\cite{Belyaev_2013} at LO and the Babayaga generator~\cite{CarloniCalame:2003yt}, \cite{Balossini:2008xr} at LO and at NLO were used to generate the kinematics of annihilation events. For certain samples, the photons of the final state were  
plugged event-by-event in GEANT4 simulations, choosing a production vertex corresponding to a location in the target reached by a positron of the incoming beam. The positron was not tracked any longer, thus emulating its annihilation in flight, and the photons were propagated through the detectors like any other primary or secondary simulated particle.

A summary of the main features of the Monte Carlo samples is reported Tab. \ref{Tab.:Samples}. 

\begin{table}[!ht]
\caption{Monte Carlo samples used in the analysis. Each sample had a size of $10^6$ events.}
\begin{center}
\begin{tabular}{ccc}
Generator & Process & Approx.\\
\hline
GEANT4&  $e^+$ Bremsstrahlung     & LO \\
      &   $e^+e^-\rightarrow \gamma\gamma$ & LO \\
      &   $e^+e^-\rightarrow e^+e^-$       & LO \\
      \hline
CalcHEP & $e^+e^-\rightarrow\gamma\gamma$ & LO  \\
\hline
CalcHEP & $e^+e^-\rightarrow\gamma\gamma$ & LO \\
+ GEANT4  & (1 $e^+$/bunch) & \\
\hline
CalcHEP &  $e^+e^-\rightarrow\gamma\gamma$ & LO \\
+ GEANT4  &  ($25000~e^+$/bunch)  &\\
\hline
Babayaga & $e^+e^-\rightarrow\gamma\gamma$ & LO\\
\hline
Babayaga & $e^+e^-\rightarrow\gamma\gamma(\gamma)$ & NLO\\
\end{tabular}
\end{center}
\label{Tab.:Samples}
\end{table}

\subsection{Event and photon selection}
\label{Sec.:EventSelection}
An event pre-selection was obtained considering only bunches where the number of POT measured by the active target was less than $\pm$5$\sigma$ away from the average. This condition rejects not interesting events, such as cosmic triggers and empty bunches, bunches with an anomalous high multiplicity, and any accidental mis-measurement of the number of POT by the target. 

Several requirements were applied to the ECAL clusters in order to obtain a clean sample of annihilation photons.
The cluster quality cuts were: cluster position offset from the crystal seed position smaller than 20 mm,
standard deviation of the $x$ and $y$ coordinates of hits in the cluster greater than 1 mm,   
standard deviation of the arrival time of hits in the cluster lower than 3~ns,  and 
 linear correlation between X and Y coordinates of cluster hits weighted by energy lower than 0.99.
In addition, an isolation cut was also applied, rejecting all clusters in time within 10 ns with a second cluster closer than 200 mm. These criteria reduced the number of clusters considered in the analysis by 50\% in average, while clusters from genuine annihilation photons were accepted with an efficiency of 90\%. 
The cluster quality cuts reduce significantly the systematic errors on the two-photon yields (see Sec. \ref{Sec.:Yield}).

\subsection{$e^+e^-\rightarrow\gamma\gamma$ kinematics}
\label{Sec.:2gKinematic}
In the Born approximation, the two-photon kinematics of the e$^+$e$^-$ annihilation process is 
highly constrained, implying the following relations \footnote{
The polar angle $\theta$ is defined as the angle between the 
photon direction and the z axis of the PADME
reference frame, which corresponds to the direction of the incoming positron beam.
The azimuthal angle  $\phi$ is the angle between the direction of a photon projected in the plane perpendicular to the beam and the horizontal x axis. 
The y axis is vertical and pointing upwards.}:
\begin{itemize}
\item 
the sum of the energies of the two photons is equal to the beam energy: $E_{\gamma_1}+E_{\gamma_2}=E_{beam}$;
\item
the photon momenta are back-to-back in the transverse plane: $\phi_{\gamma_1}-\phi_{\gamma_2}=\pi$;
\item
for each photon the polar angle is a function of the energy: $E_{\gamma_i}=f(\theta_{\gamma_i})$;
\item
The Center-of-Gravity (CoG) of the interaction in the transverse plane 
\begin{equation}
CoG_{x(y)} = \frac{x(y)_{\gamma_1} E_{\gamma_1} + x(y)_{\gamma_2} E_{\gamma_2}}{E_{\gamma_1}+E_{\gamma_2}} 
\end{equation}
is null;
\item the invariant mass of the system produced along with each photon in the $e^+e^-$ annihilation,  dubbed ``squared missing mass'' and approximated with 
\begin{equation}
\hspace{-0.7cm}
M^2_{miss} = 2m_e\Big[ E_{beam}-E_{\gamma} \Big(1+\frac{E_{beam}}{2m_e}\theta^2_{\gamma}\Big)\Big], 
\end{equation}
is close to zero.
\end{itemize}

For a given beam energy, the previous relations imply that whenever the energy or the polar angle of one annihilation photon in the pair is measured, the energy and polar angle
of the other annihilation photon can be predicted.
The relations are 
used in data to calibrate the ECAL energy response 
and to verify its assembly geometry and alignment with respect to the real beam axis and target spot.
\begin{figure}[t]
\begin{center}
\includegraphics[scale=0.3]{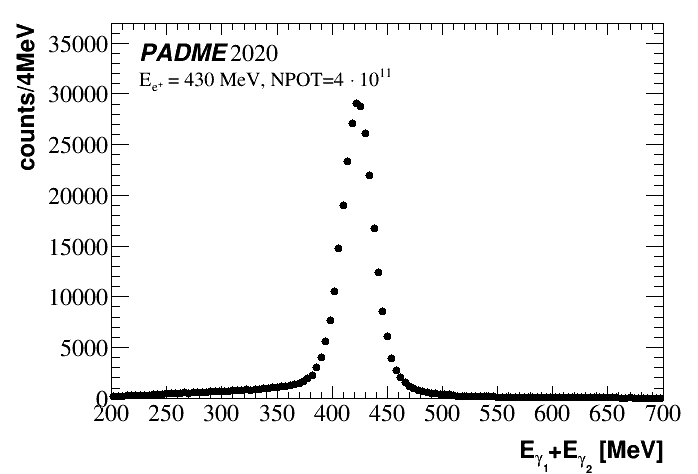}
\caption{Sum of the energies of the two photon candidates. The mean and standard deviation of the distribution core are 422.9~MeV and 14.8~MeV respectively.}
\label{Fig.:EnergySum}
\vspace{0.5cm}
\includegraphics[scale=0.3]{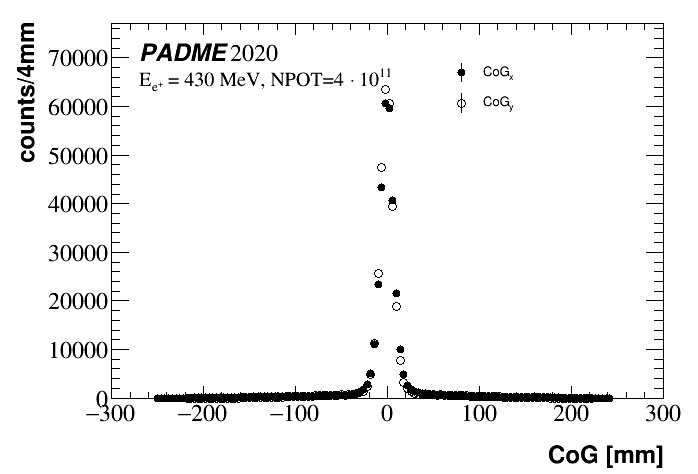}
\caption{
The Center-of-Gravity of the interaction in the $x$ and $y$ direction, reconstructed from the two photon candidates, after applying run dependent corrections. The mean and standard deviation of the distribution core are -0.38~mm and 7.61~mm in the $x$ view, -0.84~mm and 7.24~mm in the $y$ view. 
}
\label{Fig.:CoG}
\end{center}
\end{figure}

In Figure \ref{Fig.:EnergySum} 
the sum of the energy of two good quality ECAL clusters in time coincidence within 10~ns is shown for data after ECAL energy calibration (see Sec. \ref{Sec.:Reconstruction}).  

Figure \ref{Fig.:CoG} shows the distributions of 
the $x$ and $y$ coordinate of the CoG of a pair of  good quality clusters in time coincidence within 10~ns. 
These are obtained after applying event by event $x$ and $y$ shifts of the ECAL cluster positions to correct for a global offset of ECAL with respect to the beam line ($x_{ECAL}$=-3.13 mm and $y_{ECAL}$=-3.86 mm, confirmed by survey measurements) and for run dependent offsets (of the order of 1 mm) due to small changes of  the beam position and direction confirmed by the beam spot monitor provided by the active target. 

\begin{figure}[t]
\begin{center}
\includegraphics[scale=0.3]{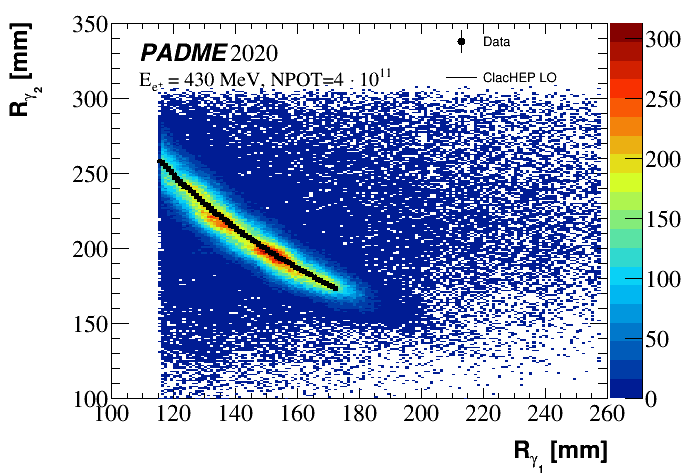}
\caption{Scatter plot between the radial positions of the leading and sub leading photons from data and CalcHEP LO.}
\label{Fig.:R1vsR2}
\end{center}
\end{figure}

In Figure \ref{Fig.:R1vsR2}  the correlation between the
radial position $R$ (which is strictly related to the polar angle $\theta$) 
of two annihilation photons is shown for the LO CalcHEP generator and for data.

\begin{figure}[htbp]
\begin{center}
\includegraphics[scale=0.3]{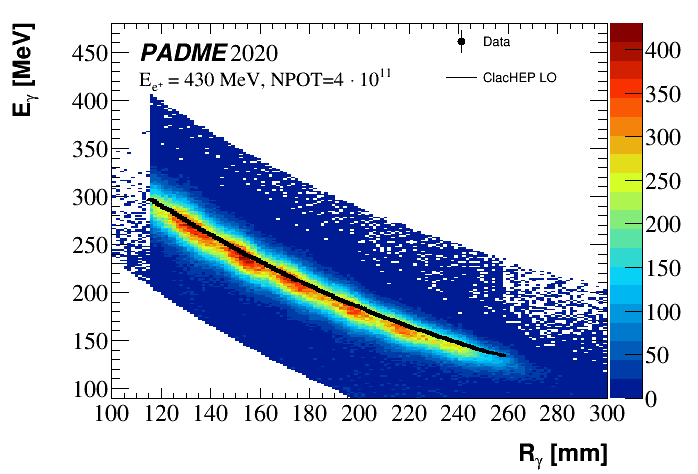}
\caption{Scatter plot between energy and radial positions of the leading and sub leading photons  from data and CalcHEP LO.}
\label{Fig.:EvsR}
\end{center}
\end{figure}

In Figure \ref{Fig.:EvsR}  the correlation between energy $E$ and radial position $R$ 
of the two annihilation photons is shown for the LO CalcHEP generator and for data after energy calibration and CoG correction.
For the simulation, the radial position of the photons in ECAL is computed as $R_{\gamma}= D \sqrt{p_x^2+p_y^2}/p_z$ in terms of the momentum components ($p_x$, $p_y$, $p_z$) and 
$D$, the sum of the distance of the ECAL front surface from the target, $D_{ECAL}$ (3470 mm), and the average e.m. shower depth  $\Delta z_{shower}$ (73 mm), as predicted by the PADME simulation. 
With this definition of the truth photon radial position the deviation between the $E(R)$ correlation in data and simulation is less then 1\% over the range of interest. 
This agreement is indicative of the accuracy of the geometrical parameters describing the ECAL crystal array in the data reconstruction and the overall ECAL alignment. For this geometry cross-check a tight selection of photon pairs reconstructed in a fiducial region of ECAL was used. 

\section{Inclusive e$^+$e$^-\rightarrow\gamma\gamma$ production cross-section}
\label{Sec.:cross-section}
The cross-section is measured as follows: 
\begin{equation}
\sigma_{e^+e^-\rightarrow \gamma\gamma} = \frac{N_{e^+e^-\rightarrow \gamma\gamma}}{N_{POT} \cdot n_{e/S} \cdot  A_g \cdot A_{mig} \cdot \epsilon_{e^+e^-\rightarrow \gamma\gamma}} 
\label{Eq.:xSec}
\end{equation}
where: 
\begin{itemize}
\item
$N_{e^+e^-\rightarrow \gamma\gamma}$ is the signal yield; 
\item
$N_{POT}=(3.97\pm 0.16)\times $10$^{11}$ is the total number of positrons on target as measured by the
active diamond target. It corresponds to the integral 
of the number of POT estimated event by event by the active target;
\item
$n_{e/S}=\rho N_A d \ Z/A$ = 0.01025 $\pm$ 0.00038 b$^{-1}$ is the electron surface density of the 
target,
where $\rm \rho = 3.515 \pm 0.015~g/cm^3$ is the diamond density, 
$Z$ and $A$ are the carbon atomic number and weight, respectively, and $d=(0.0969\pm 0.0036)$~mm is 
the estimated average diamond target thickness;
\item 
A$_g$ is the acceptance of the selection, dictated by the ECAL geometry;
\item
A$_{mig}$ is a correction for event migration across the border of the acceptance region; 
\item
$\epsilon_{e^+e^-\rightarrow \gamma\gamma}$ is the combined detection, reconstruction and selection efficiency for two-photon events within the acceptance. 
\end{itemize}

\subsection{Two-photon selection}
\label{Sec.:2gSelection}
A fiducial region, defined by 
115.82~mm$< R_{\gamma_{1(2)}} <$ 258~mm, is introduced to enforce a reliable 
reconstruction of the photons. The corresponding range of polar angles for the photons is $[32.75,72.74]$~mrad. Clusters in this region are at a distance from the inner and from the outer boundaries of the calorimeter equal to at least twice the size of a BGO crystal and this ensures a good transverse shower containment and therefore a good determination of the energy and position. The values of the edges of such region are chosen  to be consistent with the two-photon kinematics. The radial position where the two photons have equal energy,  $R_{mid}=172.83$~mm ($\theta_{mid}$=48.78~mrad), is determined consistently in data and in the LO simulation at generator level. It is used to define the inner and outer ring of the fiducial region for efficiency measurements and yield determination based on a single photon selection.

The event selection applied for the measurement of the inclusive $e^+e^-\rightarrow \gamma\gamma$ cross-section requires two photons matching the following criteria: 
\begin{enumerate}
\item time coincidence $|t_{\gamma_1}-t_{\gamma_2}|<10$ ns;
\item photon energy $E_{\gamma_{1(2)}} >$ 90 MeV;
\item consistency of energy and polar angle for each photon $|E_{\gamma_i}-f(\theta_i)|<100$~MeV;
\item radial position of the most energetic photon ($\gamma_1$) in the fiducial region : 115.82 mm $< R_{\gamma_1} <$  258 mm. 
\end{enumerate}
The fiducial region constraint, although applied explicitly only to the leading photon, in practice holds also for the second photon (for the large majority of two-photon events, where limited radiative effects are seen) and it defines an energy threshold for both photons well above the lower limit explicitly enforced by the selection. Indeed, this requirement is the only criterion defining the acceptance of the event selection. The remaining requirements are very loose compared to the detector resolution.    
Therefore, acceptance and efficiency are not strongly dependent on detailed features of the detector which could be difficult to simulate at the $\%$ level as required by the statistical precision of the analysis. 

\subsection{Two-photon annihilation yield}
\label{Sec.:Yield}
\begin{figure}[t]
\begin{center}
\includegraphics[scale=0.32]{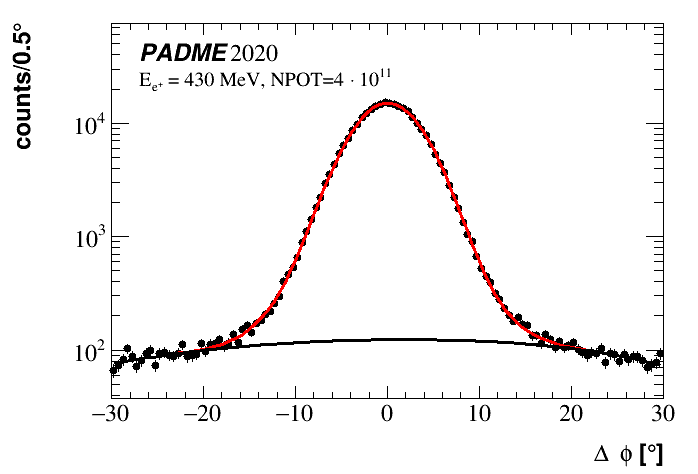}
\caption{Deviation from 180$^\circ$ of the azimuthal angle difference between the two annihilation photons candidates.}
\label{Fig.:DeltaPhi}
\end{center}
\end{figure}

The two-photon annihilation yield is estimated from  
the distribution of the variable:
\begin{equation}
\Delta\phi=\phi_{\gamma_1}-\phi_{\gamma_2}-\pi 
\end{equation}
which is symmetric around zero and with an almost flat background.
Fig. \ref{Fig.:DeltaPhi} shows the $\Delta\phi$ distribution over the entire data set of the analysis; the fitting function is the sum of two Gaussian distributions, for the signal, and a second order polynomial, for the background. After the subtraction of the background, that contaminates the selection by less than 10\%, the total number of annihilation events available for the cross-section measurement is 276700$\pm$560. 

The annihilation yield is also measured in eight azimuthal slices of ECAL, each 45$^\circ$ wide, by assigning each photon-pair candidate to the slice where the leading photon is reconstructed. The sub-leading photon in the pair lies in the opposite slice, or just across its borders due to resolution effects.

The annihilation yield has also been measured with a single-photon selection, looking for photons fulfilling kinematic conditions specific of annihilation photons, like $\Delta E \sim 0$ or $M^2_{miss} \sim 0$. The cross-section is still  computed according to Equation \ref{Eq.:xSec} with the efficiency being, 
in this case, the single-photon efficiency. 

\subsection{Acceptance}
\label{Sec.:Acceptance}
The acceptance is defined by the range allowed for the radial position of the leading photon using Babayaga at NLO, which treats consistently 
final states with two or three photons. It was measured as the fraction of generated events with at least two photons of energy above 90~MeV, $|E_\gamma-f(\theta_\gamma)| < 100$~MeV and the leading photon in the radial fiducial region. Photons separated at the ECAL surface by a distance smaller than the clusterization distance were merged together. A value of 0.06368$\pm$0.00025, where the error is statistical, was obtained. This is 1.6$\%$ lower than the acceptance estimated at LO. The comparison of the predictions by Babayaga and CalcHEP at LO showed consistency within 0.4\%.
Finally, the contribution to the systematic error on the cross-section measurement due to the acceptance cut was estimated by varying the inner edge of the radial fiducial region up to $\pm$0.5 mm, corresponding to the estimated error on the alignment from detector surveys. This error was found to impact the measurement by about 1.16$\%$.

\subsection{Event migration effects}
\label{Sec.:Migration}
In Eq.~\ref{Eq.:xSec} the effect of migration of events across the boundaries of the acceptance, induced by resolution, is corrected with the factor $A_{mig}$. Given the definition of acceptance discussed in Section~\ref{Sec.:Acceptance}, this consists in the effect of migration of the leading photon across the inner edge of the radial fiducial region, caused by resolution or biases on the measurement of the cluster position. The sample of $e^+e^- \rightarrow \gamma\gamma$ events generated with CalcHEP at the LO approximation was simulated ignoring dead channels in the calorimeter and any source of physics or beam related background. As a consequence, the distributions of energies and positions of the photons reconstructed with the same algorithm applied in data are representative of the true distributions convoluted with resolution effects, while the inefficiency is negligible.   The correction $A_{mig}$ was computed as the ratio between 
the number of events with the leading photon reconstructed at a radius $R\ge R_{min}$ and 
the number of events with the leading photon satisfying the same condition at generator level. 
The result is $A_{mig} = 0.996\pm 0.003$, where the systematic error reported is a conservative estimate of variations induced by an imperfect knowledge of the inner border of the acceptance region or a mismatch of the profile of the reconstructed radial position between data and simulation due to mismodeling of the resolution. 




\subsection{Reconstruction efficiency}
The overall single photon efficiency $\epsilon_\gamma$ was measured in data with a Tag-and-Probe technique, exploiting the closed kinematics of two-photon annihilation events. 

In particular, for annihilation photons, named tags, the 
variable $\Delta E_{tag} = E_{tag}-f(\theta_{tag})$, where $ E_{tag}$ is the measured energy and $f(\theta_{tag})$ the expected energy, must be approximately zero. This feature can be used to identify tag candidates. Each tag allows inferring the existence of a second photon, named probe, with opposite azimuthal angle $\phi_{probe}=\phi_{tag}+\pi$, energy given by  $E_{probe}$=$E_{beam}$-$E_{tag}$ and  $\Delta E_{probe}= E_{probe}-f(\theta_{probe})\approx 0$. 
If a cluster passing all photon selection requirements of the analysis has features matching those of the probe it defines a ``matched probe'', i.e. it counts in the number of efficiently reconstructed probes. 
Un-matched probes,  instead,  correspond to an inefficiency for a photon at $\phi=\phi_{probe}$ and $R=R_{probe}$ (or $E=E_{probe}$).

The efficiency $\epsilon_\gamma$ has been evaluated in 16 bins,  corresponding to eight azimuthal angle slices (each 45$^\circ$ wide) times two radial regions, the inner one going from $R_{min}$ to $R_{mid}$, and the outer one from $R_{mid}$ to $R_{max}$. 

The efficiency $\epsilon_\gamma(i,j)$ in a generic bin, identified by the index $i$, ranging over the eight azimuthal slices, and the index $j$, ranging over the two radial bins,  
is estimated as the fraction of probes, predicted in that bin by tags reconstructed in the opposite radial and azimuthal bin, which are actually matched
by reconstructed (and selected) photons, i.e.
\begin{equation}
\epsilon_\gamma(i,j)= N_{matched-probes}(i,j)/N_{tag}(i',j').
\label{eq:epsilon}
\end{equation}

\begin{figure}[t]
\begin{center}
\includegraphics[scale=0.3]{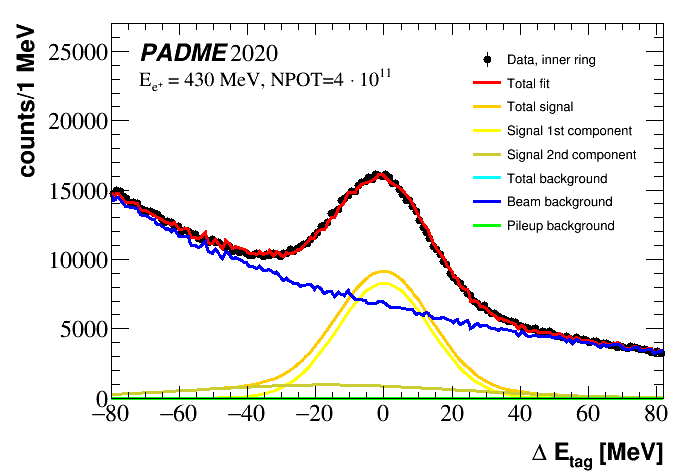}
\caption{Distribution of the $\Delta E_{tag}$ distribution for all photons in the inner ring of the radial fiducial region passing the tag selection. 
}
\label{Fig.:TagInner}
\end{center}
\end{figure}

The number of tags in a given bin, $N_{tag}(i',j')$,  
is obtained from a 
fit to the $\Delta E_{tag}$ distribution of all photons passing the following tag selection:
\begin{enumerate}
    \item $E_{tag}> 90$~MeV
    \item $|\Delta E_{tag}|<100$ MeV
\end{enumerate}
The fitting model is the sum of two Gaussian distributions, for the signal, with total yield giving N$_{tag}$, and two background templates. All signal and background components have fixed shape and floating amplitudes. 
The background originates from different sources, physics background from in-time  interactions and beam related background. 
The shape of the first component is extracted from a Monte Carlo sample with pile-up (mostly Bremsstrahlung), which is representatives of interactions in the target. The shape of the second component is extracted from no-target data, which are representatives of interactions in the detector and beam line materials (except the target and its support). The relative fraction of the two background components in the data has a strong dependence on azimuth and radial position.
The $\Delta E_{tag}$ distribution of all photons in the inner radial region $R_{min}<R_{tag}<R_{mid}$ passing the tag selection is shown in Fig. \ref{Fig.:TagInner}. 

\begin{figure}[t]
\begin{center}
\includegraphics[scale=0.3]{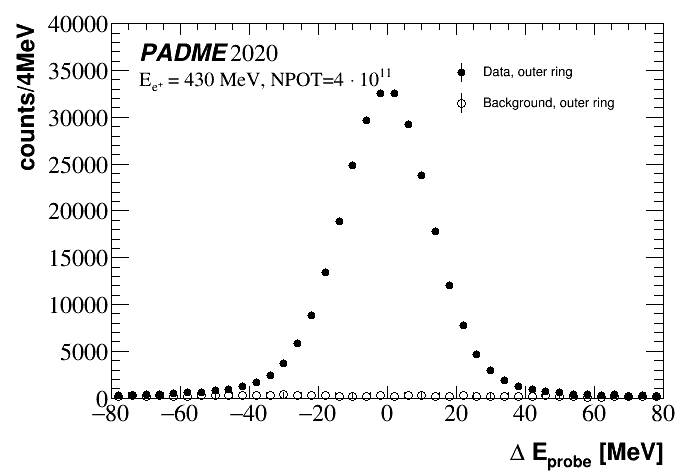}
\caption{Distribution of  $\Delta E'_{probe}$  for photons in the
outer ring of the radial fiducial region passing the matched-probe selection. The contribution of background is represented by the scaled no-target data. 
}
\label{Fig.:ProbeOuter}
\end{center}
\end{figure}

The number of matched probes in a given bin, $N_{matched-probes}(i,j)$ 
is estimated from the distribution of 
$\Delta E'_{probe}=E_{probe}-E_{beam}+f(\theta_{tag})$ for all pairs consisting of a tag-photon in the opposite bin and matched-probe photon passing the following selection criteria:
\begin{enumerate}
    \item $\phi_{tag}\in$ bin $i'$ and $R_{tag} \in$ bin $j'$;
    \item $|\phi_{tag}+\pi - \phi_{probe}|<25^\circ$ 
    \item $|t_{probe}-t_{tag}| < 7$~nsec and $E_{probe}>$90 MeV
    \item $|\Delta E_{probe}|<100$~MeV and $|\Delta E'_{probe}|<100$~MeV.
\end{enumerate}
If more than one matched probe is found for a given tag, the photon with the minimum value of $(\Delta E'_{probe})^2+(\Delta E_{probe})^2$ is selected.
The background in the $\Delta E'_{probe}$ distribution is estimated with a template, obtained in no-target data, of fixed shape and amplitude constrained by matching the yield of the scaled no-target template to target data in the left side-band, $[-150, -90]$ MeV.
The background of the matched probe distribution is quite small, therefore the contribution from  pile-up is neglected.
Fig. \ref{Fig.:ProbeOuter} shows the $\Delta E'_{probe}$ distribution and the scaled no-target data of all photons within the outer radial region $R_{mid}<R_{tag}<R_{max}$ 
passing the matched probe selection. The yield of matched probes is estimated as the integral of the $\Delta E'_{probe}$ distribution, in the range $[\mu-3\sigma, \mu+3\sigma]$, where $\mu$ and $\sigma$ are the parameters obtained from a Gaussian fit of the core of the distribution, subtracted by the background yield evaluated in the same range.

\begin{figure}[t]
\begin{center}
\includegraphics[scale=0.3]{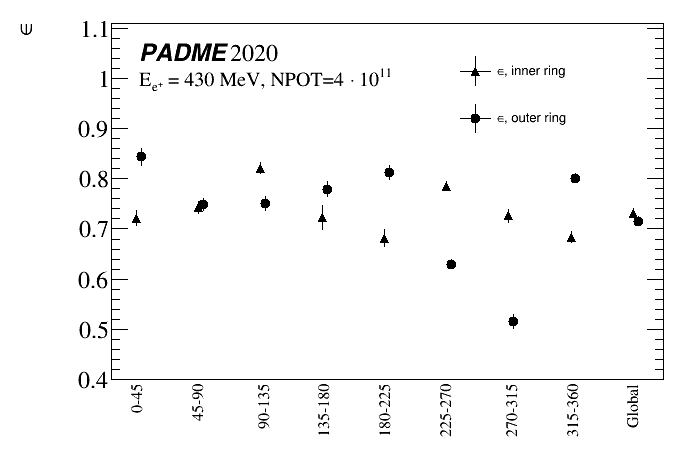}
\caption{Tag-and-Probe efficiency for 8 ECAL slices and 2 radial regions and the global inner and outer efficiency.}
\label{Fig.:TandPEff}
\vspace{1cm}
\includegraphics[scale=0.3]{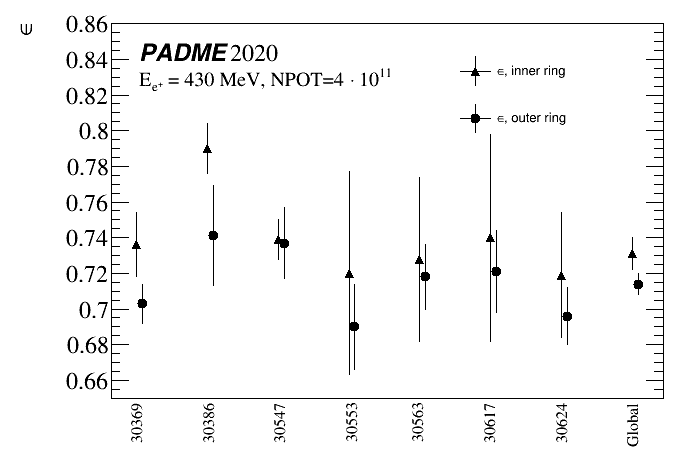}
\caption{Single photon
Tag-and-Probe efficiency for the 7 runs and for all runs in the inner and outer radial region.}
\label{Fig.:TandPEffRuns}
\end{center}
\end{figure}

The statistical error associated to the tag and matched probe yields are propagated to the efficiency. The photon detection and selection efficiency $\epsilon_\gamma$ measured in the 16 independent regions of ECAL is shown in Fig. \ref{Fig.:TandPEff}, where only statistical errors are considered. The differences from bin to bin are the result of local defects (three inefficient photo-multipliers), asymmetric geometrical acceptance of the PADME detector and inert materials, and non-uniform background rate in ECAL. Fig. \ref{Fig.:TandPEffRuns} shows the single photon efficiency, averaged over $\phi$ and $R$, estimated separately for each runs in the inner and outer ring of the fiducial region.

The Tag-and-Probe technique was validated in MC, by simulating ECAL with dead crystals and applying two closure tests. First, the Tag-and-Probe efficiency measured in simulation was compared with the truth efficiency after selection cuts. Second, the annihilation cross-section measured in simulation, with the same procedure applied in data,  was compared to the truth cross-section.
In both tests the agreement was 
well below 1\%. 

Due to the correlation between photon energy and radial position exploited in Tag-and-Probe technique, these efficiency measurements are well suited for annihilation photons and in general do not apply to photons produced in other physics processes.

The event efficiency $\epsilon_{e^+e^-\rightarrow \gamma\gamma}$ is given by the product  $\epsilon_{\gamma_1}\epsilon_{\gamma_2}$. 
The overall single photon efficiency obtained considering tags and matched probes from all slices together is measured to be 0.731$\pm$0.009,  for photons in the inner ring,  and 0.714$\pm$0.006, for the photons in the outer ring. 
These values are dominated by the tight selection requirements. 

\subsection{Systematic uncertainties}
\label{Sec.:Syst}
Several contributions to the systematic uncertainty 
on the cross-section measurement were evaluated.

\begin{figure}[t]
\begin{center}
\includegraphics[scale=0.3]{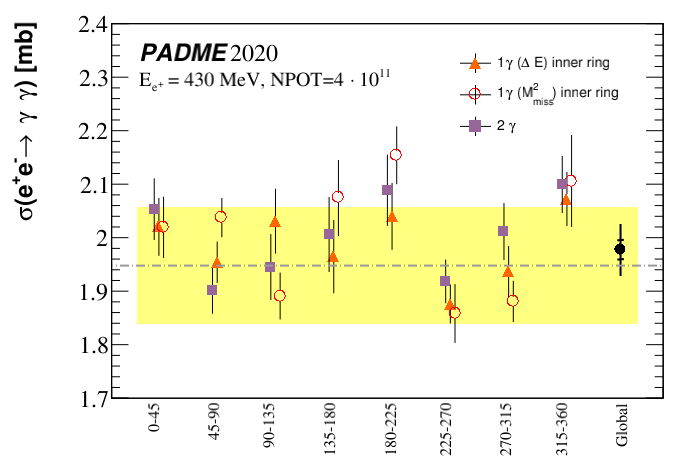}
\caption{Cross-section for the process $e^+e^-\rightarrow \gamma\gamma$
measured in different ECAL azimuthal angle slice (degrees). The error bars represent the statistical error on the measurements. The dotted line represents the QED prediction at NLO estimated with the Babayaga generator and the yellow band corresponds to the uncertainty on the number of collisions recorded.} 
\label{Fig.:SplitPhi}
\end{center}
\end{figure}

The signal yield is not expected to be uniform  
with the azimuthal angle due to local detector defects
and to the presence of the dipole magnet and vacuum vessel. Moreover, in ECAL the halo of the beam background is offset with respect to beam axis. As a conseguence, the background subtracted in the single-photon selection varies for a factor up to 3 in the inner ring of the fiducial region and up to 7 in the outer ring.
The local determination of the photon efficiency by a data-driven method 
allows compensating for these effects. However, residual systematic biases are possible due to the large differences between the average efficiency observed from bin to bin. 
The data sample is split in eight azimuthal slices and  independent cross-section measurements, one per slice, were performed. The value of the acceptance of a single slice is computed as ${A_g}_i=A_g/8$ and the event efficiency as the product of the local efficiencies relevant for the slice $\epsilon_\gamma(i,j\!\!=\!\!1) \epsilon_\gamma(i',j\!\!=\!\!2)$, where $i$ is the slice index ($i'$ the index of the opposite slice) and $j\!=$1(2) corresponds to the inner(outer) bin in the slice.
The variance of these measurements is found to exceed the expected statistical fluctuations around the weighted average, pointing to the presence of systematic local biases. 
A systematic uncertainty is therefore assigned to the cross-section measurement, to account for biases in the efficiency due to local defects and uneven background distribution, estimated as:
\begin{equation}
\sigma_{{eff},\phi} = \sqrt{[{\tt RMS}(\sigma_i)]^2 - \delta^2_{stat}(<\!\!\sigma\!\!>)}
\label{Eq.:SpliSample}
\end{equation} 
In equation \ref{Eq.:SpliSample}, $\sigma_i$ refers to the cross-section measured in slice $i$, $<\!\!\sigma\!\!>$ is the weighted average of the cross-sections and $\delta_{stat}(<\!\!\sigma\!\!>)$ is its statistical error. 
This procedure is applied to the measurements, shown in Fig. \ref{Fig.:SplitPhi},  obtained with three methods: using the two-photon selection, and using the single-photon selection, in the inner ring of the fiducial region, based on  the $\Delta E$ or the $M^2_{miss}$ distribution. The average of the three estimates is used as systematic uncertainty on the slice  dis-uniformity. 

\begin{figure}[htbp]
\begin{center}
\includegraphics[scale=0.3]{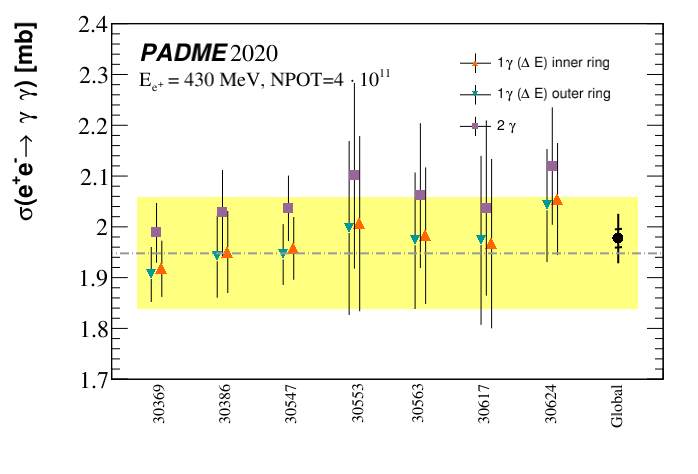}
\caption{Cross-section for the process $e^+e^-\rightarrow \gamma\gamma$
measured in each run separately. The error bars represent the statistical error on the measurements. 
The dotted line shows the QED prediction at NLO estimated with the Babayaga generator and the yellow band corresponds to the uncertainty on the number of collisions recorded.}
\label{Fig.:SplitRun}
\end{center}
\end{figure}

The same procedure was applied splitting the data sample in independent sub-samples corresponding to different runs (see table \ref{Tab.:Runs}). The cross-section was measured in each sub-sample, collecting  event candidates from all sectors together, but using a determination of the efficiency estimated for each specific run both for the inner ring (i.e. the set of eight inner bins) and for the outer. Deviations in excess of 
the statistical fluctuations might be indicative of biases in the efficiency related to the different background conditions determined by the varying density in time of positrons in the beam bunches. 
Fig. \ref{Fig.:SplitRun} shows the measurements for each run together with the measurements with all runs. Since the fluctuations are fully consistent with the expected statistical error, no systematic uncertainty is assigned.

The background modelling is a potential source of systematic uncertainty. It affects the determination of the efficiency, through the counting of the number of tag photons as well as the yield extraction in the single-photon selection. Since the background contaminates the  $\Delta E=E-f(\theta)$ and $M^2_{miss}$ distributions with a different shape, the procedure applied to assess the resulting systematic error on the cross-section measurement was to compare the cross-sections obtained with  different methods: two-photon selection, single-photon selection with the yield extracted from the $\Delta E$ distribution, and with the yield extracted from the $M^2_{miss}$ distribution. 
Only photons of the inner radial region were considered, because outer and inner signal photons are related to the same events and their distributions are just a different representations of the relationship between photon energy and polar angle. 
The set of the most energetic photons of the two-photon selection is almost a complete overlap with the set of photons of the inner radial region.  Therefore, the differences among these three measurements
have only a systematic origin, to be ascribed, as anticipated, to the  background modelling.
The systematic error is estimated as the root mean square of the
three values multiplied by $\sqrt{3}$, in order to compensate the suppression factor that would hold in case of  statistical independence of the measurements. This fit-variant procedure gives an error of 0.009 mb.

In addition, the difference between the cross-section obtained in the single-photon selection
applied in the inner and outer rings was considered. The two measurements are based
on consistent, but experimentally different, definitions of the fiducial region.
Therefore, the difference between the two measurements, based on  $\Delta E_{in}$ and $\Delta E_{out}$, can be considered as an indication of a systematic uncertainty related to  the acceptance. 
This has been assessed as the root mean square of the two values multiplied by $\sqrt{2}$
because exactly the same events are involved. The procedure was repeated with the cross-sections measured by single-photon selections in the inner and outer fiducial region based on $M^2_{miss}$. 
The average of the two estimates of this effect, equal to 0.024~mb,  
is assumed as a contribution to the systematic error due to the acceptance. 
In addition, the uncertainty on the knowledge of the boundary of the fiducial region (discussed in Section \ref{Sec.:Acceptance}) impacts with an additional 0.023~mb due to the resulting error on the acceptance and with a 0.006~mb due to the error on the correction $A_{mig}$ accounting for resolution effects. These two uncertainties are treated as fully correlated and when combined with the other sources of error lead to a total systematic uncertainty for the acceptance and event migration equal to  0.037~mb.

Other contributions to the error on the cross section measurement come from the uncertainty on the total number of collisions. This originates from the error on the number of positrons on target, $N_{POT}$, and from the error on the electron surface density in the target, $n_{e/S}$. 

\begin{table}[t]
\centering
\caption{Systematic uncertainties on the inclusive annihilation cross-section measurement.
\label{Tab.:xSecMeasurementAndSys}}
\vspace{0.3cm}
\begin{tabular}{c|c}
\hline 
Source &  systematic error  \\
\hline
Azimuthal angle dis-uniformity   & $ 0.024 \ \rm{mb}$\\
Beam conditions  & - \\
Background modelling  & $ 0.009 \ \rm{mb}$\\
Acceptance and resolution& $ 0.037 \ \rm{mb}$\\ \hline
Total  &  0.045 mb\\ \hline \hline 
Number of POT  & $0.079 \ \rm{mb}$ \\
Target electron surface density  & 0.073 mb \\ \hline 
Total  & $0.110\ \rm{mb}$ \\ \hline
\end{tabular}
\end{table} 
The relative uncertainty on $N_{POT}$ (see Section \ref{Sec.:cross-section}) is 4\% and is dominated by the uncertainty on the absolute charge calibration of the active target response. The absolute scale of the charge response was cross-calibrated with a lead-glass Cherenkov calorimeter working in full-containment mode. Intensities from 5$\times 10^3$ up to 35$\times 10^3$ 430~MeV positrons per bunch were considered. In the data-taking campaign of Fall 2020 the beam spot was focused on a few target strips in both directions and an ad-hoc non-linear calibration was necessary to determine the number of POT, reaching a precision of 4$\%$. This uncertainty was derived by comparing in several runs the measurement of the number of POT per bunch from the diamond target, averaged over the run, with measurements of the beam intensity per bunch performed at the start and at the end of the run with the calibration calorimeter. 

The uncertainty on $n_{e/S}$ (reported in Section \ref{Sec.:cross-section}) is dominated by the error on the diamond target thickness $\sigma_d=0.0036$~mm. 
The average thickness of the active target was measured, after assembly, using an optical profilometer with a 1~$\mu$m spatial resolution as the difference with respect to a supporting surface. Because of the roughness of the unpolished diamond surface (Ra~=~3.2~$\mu$m according to the producer) a correction needs to be applied to such measurement. This is obtained  by comparing the result of the same procedure with precision mass and surface measurements on other similar CVD diamond samples. The error accounts for the statistical uncertainty on the measurements performed with the profilometer, but is dominated by the systematic component assessed as one half of the roughness related correction.

All sources of systematic errors are summarized in Tab. \ref{Tab.:xSecMeasurementAndSys}.

\subsection{Inclusive e$^+$e$^-\rightarrow\gamma\gamma$ cross-section results}

The local variations of the photon efficiency suggest to measure the cross-section in eight independent azimuthal angle slices of ECAL exploiting the granularity of the efficiency measurement. Eventually, the measurements are statistically combined in a weighed average. 


The cross-section measurements 
from the three methods  (two-photon selection, and single-photon selection in the inner fiducial region exploiting  $\Delta E$ and  $M^2_{miss}$) 
can be considered equivalent, since they are based on the same sample of annihilation events.
Therefore, they are combined in a simple average, giving the final inclusive cross-section measurement:


\begin{equation}
\begin{array}{c}
\sigma_{e^+e^-\rightarrow \gamma\gamma} [{\rm PADME}] =1.977\pm0.018 \ (stat)\\
\pm0.045\  (syst)\pm0.110 \ (n.\ collisions)\rm{\ mb}.
\end{array}
\label{Eq.:xSecMeasurement}
\end{equation}

The systematic error is the combination of the experimental systematic uncertainties described in section \ref{Sec.:Syst}. The error coming from the uncertainty on the total number of 
collisions is quoted separately.

\begin{figure}[t]
\begin{center}
\includegraphics[scale=0.28]{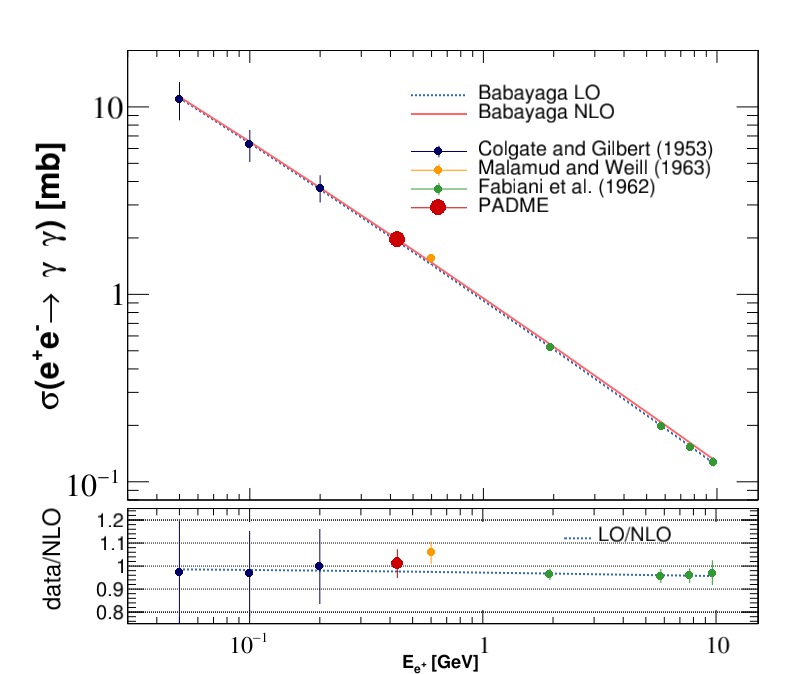}
\caption{Theory predictions, at the leading order and next-to-leading order approximation,   for the positron annihilation cross-section in flight as a function of the positron energy. The PADME measurement is superimposed along with earlier measurements. Data to theory ratios are shown in the bottom pad. 
}

\label{Fig.:cross-sections}
\end{center}
\end{figure}

The measurement is compatible with the QED prediction at NLO, estimated with the Babayaga generator~\cite{Balossini:2008xr},
\begin{equation}
\begin{array}{c}
\sigma_{e^+e^-\rightarrow \gamma\gamma} [{\rm Theory}] = 1.9478\pm 0.0005 \ (stat)\\
\pm 0.0020\  (syst)\rm{\ mb} 
\end{array}
\label{eq:theory}
\end{equation} 
for the inclusive in-flight annihilation cross section at the positron energy of $E_{e+}=432.5$~MeV. 
In Equation~\ref{eq:theory}, the statistical error comes from the statistics of the MC generation and the systematic error is a conservative estimate of higher-order corrections.
Fig. \ref{Fig.:cross-sections} shows the comparison of the  two-photon annihilation cross-section measured by PADME with the theoretical predictions and  the other measurements performed in the past at a similar energy scale.

\section{Summary}
\label{Sec.:Summary}
The inclusive annihilation cross-section of 432.5~MeV positrons with atomic electrons in Carbon leading to two-photon final states has been measured with the 
PADME experiment. The high accuracy measurement is performed using a high-granularity electromagnetic calorimeter made of BGO crystals and a pulsed beam with about 27,000 positrons per bunch. 
A Tag-and-Probe technique was applied to measure the efficiency for annihilation photons as determined by detector
defects and asymmetries in the acceptance of the apparatus, reconstruction efficiency, physics and beam related background.
The PADME result on Run II data subset is: 
\begin{equation*}
\sigma_{e^+e^-\rightarrow \gamma\gamma} = (\ 1.977\pm0.018_{stat}
\pm0.119_{syst}\ ) \rm{\ mb}
\end{equation*}
is in good agreement with NLO QED predictions. 

The PADME measurement exploits the reconstruction of the photon pair for the first time at beam energies below 1 GeV. Previous results~\cite{Colgate, Malamud} were based on the measurement of the rate of positron disappearance, which might receive contributions from beyond-the-Standard-Model processes leading to undetected final states. The PADME measurement, instead, is free from assumptions about new physics producing invisible final states. 

The experimental and analysis techniques reported here pave the way for accurate Standard Model QED measurements and searches for New Physics in positron annihilation in flight, such as MeV mass scale dark photons and axions. 

\section{Acknowledgements}
\label{Sec.:Acknowledgements}

The PADME collaboration would like to thank all LNF staff that gave support to the experiment in any phase. In particular all LINAC operators, BTF staff members and the personnel of the vacuum service of the accelerator division. All the technicians of the research division, and in particular C. Capoccia, E. Capitolo, S. Ceravolo. The collaboration wishes to acknowledge  the continuous support from the technical personnel from INFN Lecce and Dipartimento di Matematica e Fisica of Universit\`a del Salento, in particular R. Assiro, A. Miccoli, and C. Pinto. 
A special thank goes to C. Carloni Calame for the precious help in estimating the NLO cross section and kinematics using the Babayaga generator. 
This paper benefited by a support from TA3-LNF as part of the STRONG-2020 project -- EU Grant Agreement 824093.


\bibliographystyle{elsarticle-num} 
\bibliography{bib2_Padme.bib}






\end{document}